# Phase transitions via selective elemental vacancy engineering in complex oxide thin films

Sang A Lee[1,2], Hoidong Jeong[1], Sungmin Woo[1], Jae-Yeol Hwang[3], Si-Young Choi[4], Sung-Dae Kim[4], Minseok Choi[4], Seulki Roh[1], Hosung Yu[5], Jungseek Hwang[1], Sung Wng Kim[3,5], and Woo Seok Choi[1*]

[1]Department of Physics, Sungkyunkwan University, Suwon, 16419, Korea

[2]Insitute of Basic Science, Sungkyunkwan University, Suwon, 16419, Korea

[3]Center for Integrated Nanostructure Physics, Institute for Basic Science (IBS) Sungkyunkwan University, Suwon 16419, Korea

[4]Materials Modeling and Characterization Department, Korea Institute of Materials Science, Changwon 51508, Korea

[5]Department of Energy Sciences, Sungkyunkwan University, Suwon 16419, Korea

*e-mail: choiws@skku.edu.

**Defect engineering has brought about a unique level of control for Si-based semiconductors, leading to the optimization of various opto-electronic properties and devices. With regard to perovskite transition metal oxides, oxygen vacancies have been a key ingredient in defect engineering, as they play a central role in determining the crystal field and consequent electronic structure, leading to important electronic and magnetic phase transitions. Therefore, experimental approaches toward understanding the role of defects in complex oxides have been largely limited to controlling oxygen vacancies. In this study, we report on the selective formation of different types of elemental vacancies and their individual roles in determining the atomic and electronic structure of perovskite $SrTiO_3$ (STO) homoepitaxial thin films fabricated by pulsed laser epitaxy. Structural and electronic transitions have been achieved via selective**

3
**control of the Sr and oxygen vacancy concentrations, respectively, indicating a decoupling between the two phase transitions. In particular, oxygen vacancies were responsible for metal-insulator transitions, but did not influence the Sr vacancy induced cubic-to-tetragonal structural transition in epitaxial STO thin film. The independent control of multiple phase transitions in complex oxides by exploiting selective vacancy engineering opens up an unprecedented opportunity toward understanding and customizing complex oxide thin films.**

**Introduction**

Crystal defects can have a large impact on the characteristics and performance of a material affecting the crystalline and electronic structure, electrical, ionic, magnetic, optical, and mechanical properties.[1-4] Therefore, in designing a material with a set of target properties, it is crucial to understand the role of elemental defects and also be able to engineer various types and/or concentrations of defects. While defect engineering in binary systems, including binary oxides, has been extensively studied,[5-7] efforts to control individual types of defects in more complex systems, *e.g.*, perovskite oxides ($ABO_3$), are uncommon.[1,4] This is mainly due to the difficulty in systematically controlling distinctive defects in complex oxides, as structures can adopt various types and degrees of elemental vacancies and vacancy clusters. The ability to control and customize defects in complex oxides, therefore, is essential toward manipulating the versatile physical properties of transition metal oxides.[8]

Perovskite strontium titanate ($SrTiO_3$, hereafter abbreviated as STO) can be considered as a model system in the study of defect-related physics in complex oxides, since STO exhibits various intriguing physical property changes including superconductivity-metal-insulator and ferroelectric-paraelectric transitions.[9-11] STO has a perovskite cubic structure with the lattice constant of 3.905 Å and is a paraelectric band insulator with a wide band-gap of ~3.2 eV and a large dielectric constant (~$10^4$ at low temperatures). With a small amount of oxygen vacancies, STO becomes an *n*-type conductor and even a superconductor at low temperature ($T_c$ = ~0.2 K) depending on the oxygen vacancy concentration.[10] On the other hand, STO can also become ferroelectric-like through the formation of SrO vacancies.[12-14] Because the crystal and electronic structures of STO can be largely modified via elemental vacancies, defect engineering in STO can provide exceptional insight toward understanding the complex oxides with various phase transitions.

As with many transition metal oxides, oxygen vacancies have been a center of interest among traditional STO defect studies, since they are thought to affect both the structural and electronic properties of the material.[15-17] For example, oxygen vacancies induce metallic behavior through the donation of two electrons which occupy the empty Ti $3d$ band. It is also commonly believed that charged oxygen vacancies expand the crystal lattice of complex oxide materials by increasing Coulomb repulsion between neighboring transition metal ions. Furthermore, oxygen vacancies are technologically important for STO, since they trap charge carriers which are responsible for undesirable leakage currents in microelectronic devices such as the dynamic random access memory capacitors.[18,19] On the other hand, cation vacancies also influence the physical properties of STO. For example, cation off-stoichiometry (due to Sr and/or Ti vacancies) results in lattice expansion, decreases electric conductivity, and induces ferroelectricity in combination with epitaxial strain.[12,13,20-23] A significant amount of Ti vacancies gives rise to Ruddlesden-Popper-type planar faults, with the collapse of the perovskite structure.[23-25] While both oxygen and cation vacancies play critical roles in STO, systematic studies with regard to individual vacancy types and their coupling to distinctive physical properties have been largely unexplored in STO.

In this study, we provide a way to selectively control the formation of elemental vacancies in STO thin films. In order to minimize the effect of epitaxial strain, we employed homoepitaxial STO thin films on STO substrates. Figure 1 summarizes our study of elemental vacancy engineering and resultant phase transitions. In particular, we demonstrate individual control over strontium and oxygen vacancies via pulsed laser epitaxy (PLE, also known as pulsed laser deposition), and reveal the independent role of each vacancy type. While strontium vacancies are responsible for structural phase transitions in epitaxial STO thin films, oxygen vacancies are not mainly related to lattice expansions but induce electronic metal-insulator transitions.

## Results and discussion

For individual control of elemental vacancies, we modified different growth parameters using PLE. PLE is a well-established technique for the growth of perovskite oxide thin films and is advantageous since it transfers the target material to the substrate with high crystalline quality. However, as an increasing number of recent studies reveal,[23,26-28] the stoichiometry of the resultant complex oxide thin film tend to deviate from that of the target material. Furthermore, here we show that cation and oxygen vacancies can be individually controlled via careful design of the PLE growth.

Figure 2 shows the structural and optical properties of homoepitaxial STO thin films grown at different $P(O_2)$. As shown in the XRD $\theta$-$2\theta$ scan of Fig. 2(a), the STO film peak is separated from the substrate peak to a lower $2\theta$ angle, indicating that the $c$-axis lattice constant of the film is larger than that of the substrate. Although the actual crystal structure is different from that of the stoichiometric bulk material, all STO films were epitaxially grown with homogeneous flatness and good out-of-plane orientations, judging from the well-defined Kiessig and Pendellösung fringes. The atomic force microscopy topographical image of the film (see Supplementary Fig. S1) indicated that the one unit cell height step-and-terrace structure of the substrate was preserved. The full-width-at-half-maximum values of the $\omega$-scan peaks around the (002) STO Bragg diffraction shown in Figs. 2(b) and 2(c) for the substrate and the film grown at $P(O_2) = 10^{-4}$ Torr, respectively, also suggest good crystallinity of the homoepitaxial thin film. Figure 2(d) presents the $\varphi$-scans of the film grown at $P(O_2) = 10^{-4}$ Torr and the substrate around the (024) STO Bragg diffraction. Both the film and substrate reveal a four-fold symmetry and epitaxial relationship of $[001]_{film}$ // $[001]_{substrate}$. The XRD results indicate that the tetragonal STO thin films were homoepitaxially grown on cubic STO substrates with the elongated $c$-axis lattice constant due to the interplay between lattice expansion and epitaxial strain.

As shown in Fig. 2(a), the $c$-axis lattice constant (or tetragonality) of the STO homoepitaxial thin films increased with decreasing $P(O_2)$. At first sight, this observation appears to be due to oxygen vacancies induced during low $P(O_2)$ film growth and consistent with the common belief that charged oxygen vacancies expand the perovskite crystal lattice. However, upon performing detailed experiments under varying $P(O_2)$, we determined that the lattice expansion does not show a simple linear relationship with $P(O_2)$, but exhibits a wide distribution even among samples deposited at the same $P(O_2)$, as shown in Fig. 3(a). While the overall trend of the $c$-axis lattice constant shows larger expansion for films grown at lower $P(O_2)$, the rate of expansion is scattered from 0 to more than 1%. We further note that oxygen vacancies alone cannot explain the observed degree of expansion (~1%), even with a large amount present. For example, bulk $SrTiO_{2.72}$ is known to have only a 0.3% expanded lattice.[17] Such a finding suggests that oxygen vacancy, which is controlled by $P(O_2)$, may not be the main source of lattice expansion in STO.

Among other growth parameters, we note that laser energy clearly induces a systematic modification to crystal structures. Figure 3(b) shows the $c$-axis lattice expansion as a function of the laser energy used. Compared to the $P(O_2)$ dependence, the $c$-axis lattice expands in a systematic fashion with increasing laser energy for any given $P(O_2)$ ($10^{-1}$ [red axis] or $10^{-6}$ [purple axis] Torr). It has been previously reported that cation stoichiometry (Sr/Ti ratio) can be modified by changing the laser energy applied during PLD.[20,26,29,30] Since most measurement tools such as Rutherford back scattering spectrometry, energy dispersive x-ray spectroscopy, and x-ray photoemission spectroscopy accompany large uncertainties especially in determining a few percentage of stoichiometry deviations observed in our experiments, we matched our STO thin film structures to the reference data, instead of performing separate chemical analyses. Indeed, increasing tetragonality in our thin films could be well-matched with the previous results (as well as theoretical calculations shown later), in which the lattice expansion has been understood in terms of Sr deficiency (upper $x$-axis of Fig. 3(b)). This agreement allows for Sr vacancies to be estimated in the homoepitaxial STO thin films, up to ~2%.

The interplay of laser energy and $P(O_2)$ on the cation stoichiometry of STO can be summarized with plume dynamics during PLE growth. First, laser energy irradiated onto the target induces the formation of the plume.[30] The energy of the pulsed laser transfers to the target material and determines the initial velocity of the ablated elements and corresponding plume propagation. Therefore, different elemental species can be ablated at different laser energy-dependent rates. In particular, for STO, higher laser energy ablates relatively more Ti and results in Sr-deficient films, possibly due to the bonding energy difference between the cations in the crystal lattice.[30,31] Second, ablated elemental species experience different scattering conditions as they travel toward the substrate, affecting film stoichiometry.[32] For example, during the growth of STO, the scattering of lighter Ti ions is more susceptible to changes in $P(O_2)$ (e.g., Ti ions scatter more easily at high $P(O_2)$), and thus, results in relatively Sr-deficient films at low $P(O_2)$, coinciding with our experimental results. The stronger dependence of laser energy compared to $P(O_2)$ on $c$-axis lattice expansion, and the corresponding degree of Sr vacancies, suggests that scattering with oxygen molecules (the second mechanism) is more difficult to systematically control during the energetic PLE process. $P(O_2)$ obviously changes the number of oxygen vacancies in the film, in addition to the number of cation vacancies, while the influence of laser energy on oxygen stoichiometry is rather limited. Nevertheless, we demonstrated that both parameters could be separately tuned to control Sr vacancies, inducing lattice expansion in STO thin films.

We further performed theoretical calculations in order to support the observed Sr vacancy induced structural transition in epitaxial STO thin films. Figure 4 presents the relative total energy of a defect-containing supercell as a function of $c$-axis lattice expansion for different defect configurations. Four different types of vacancies, *i.e.*, Sr, Sr-O, oxygen on the TiO$_2$ layer ($O_{ab}$), and oxygen on the SrO layer ($O_c$), have been considered (as shown in the inset). The in-plane lattice constant was fixed to the bulk STO value to represent the homoepitaxial growth. Results indicated that different defect structures accommodated distinct lattice structures (tetragonal distortion). In particular, about 0.5% expansion in the out-of-plane direction is expected for defect structures containing Sr (Sr and Sr-O vacancies). On the contrary, with only oxygen vacancies, ~0.5% contraction of the $c$-axis lattice constant is expected, as opposed to expansion. Therefore, expansion of the crystal lattice is only expected for Sr-containing vacancies, which is in good agreement with our experimentally observed results. The calculated lattice expansion value for the Sr vacancies is also plotted in Fig. 3(b) as the green star, indicating a quantitative agreement between the theoretical and experimental results.

In addition to lattice expansion, a systematic color change was observed for the STO films grown at varying $P(O_2)$. As shown in Fig. 2(e), the originally transparent STO films became dark blue with decreasing $P(O_2)$. We note that the optical properties of the STO thin films are almost identical if the same $P(O_2)$ (and oxygen flow rate) is used, which is quite different from the structural properties. The observed color change could be attributed to optical absorptions at ~1.5, ~2.3 and ~2.9 eV, due to impurity bands generated by oxygen vacancies.[12,33,34] We confirmed that the change in the optical properties comes mainly from the STO thin films, since the color of the substrates did not change significantly under the growth conditions used. In particular, we have annealed the STO substrates in $P(O_2) \geq 10^{-5}$ Torr at 700 °C for 20 minutes (identical to deposition condition), but did not observe any change in the optical properties. To our surprise, the color of the thin films could also be modified by adjusting the oxygen flow rate. In particular, at $P(O_2) = 10^{-4}$ and $10^{-5}$ Torr, the oxygen flow rate could be modified by controlling both the inlet and outlet gas conductance in the vacuum chamber, while maintaining $P(O_2)$ (Fig. 2(e)). To minimize the effect of local pressure fluctuations and/or gradient in $P(O_2)$, we have monitored the pressure in three different locations within the vacuum chamber. $P(O_2)$ is the oxygen partial pressure, which is ideally the number of oxygen gas molecules inside the chamber at a specific time. On the other hand, oxygen flow rate is the number of oxygen passing by the chamber (from gas inlet to the vacuum pump) within a given time period. Therefore, in principle, the two values are independent of each other. The maximum flow rate (1.0 (0.1) sccm for $10^{-4}$ ($10^{-5}$) Torr) could be achieved by fully opening the valves (high flow, HF). A substantially lower flow rate (0.03 (0.07) sccm for $10^{-4}$ ($10^{-5}$) Torr) could be achieved by partially opening the throttle valve (low flow, LF), for the same $P(O_2)$. As shown in Fig. 2(e), STO homoepitaxial films grown under LF conditions exhibited remarkably darker colors compared to films grown at HF conditions, although the actual quantity of oxygen molecules inside the vacuum chamber was equivalent. This observation implies that HF of oxygen gas supplies more oxygen to the STO thin films, dynamically compensating for more oxygen vacancies. This is because there are actually more oxygen gas molecules within a given time period (e.g., growth time), which can interact either with the plasma plume or the STO thin films.

The use of oxygen flow rates in determining the concentration of oxygen vacancies yields important insight toward controlling elemental vacancies and engineering phase transitions in complex oxides. As the oxygen flow rate does not affect plume dynamics significantly, it leads to a selective control of oxygen vacancies without largely influencing cation stoichiometry. Note that such an approach is not plausible via conventional $P(O_2)$ control. Figure 5 reveals the optical properties and crystal structure of STO thin films grown at $P(O_2) = 10^{-4}$ Torr. Figure 5(a) presents the transmittance and absorption spectra (inset) indicating that samples grown under HF are more transparent compared to sample grown under LF of oxygen. While the films grown at HF show signs of oxygen vacancies (absorptions at ~1.5 and ~2.9 eV) compared to stoichiometric STO substrates, the films are still largely transparent. However, LF grown samples exhibit a large absorption at lower photon energies due to the Drude contribution. This indicates that the sample has become metallic due to free charges induced by the oxygen vacancies (see Supplementary Fig. S2). Figure 5(b) shows that HF grown films (low oxygen vacancy concentrations) yield more expanded $c$-axis lattice constants compared to the LF grown film (high oxygen vacancy concentration). However, the actual $c$-axis lattice constant difference between the two samples is rather small (less than 0.4%), indicating that the changes in cation stoichiometry are subtle. We also performed annealing (at 600 °C for an hour in air) experiment for the STO thin film grown at $P(O_2) = 10^{-6}$ Torr. After annealing, the originally dark blue film becomes transparent, indicating that the oxygen vacancies are presumably been compensated. The lattice constant of the film, however, did not retract to the original bulk value after the annealing. Instead, it increased slightly, again suggesting that introduction of oxygen vacancies does not expand the lattice constant (see Supplementary Fig. S3). Hence, our results strongly support the notion that oxygen vacancies can indeed be independently controlled during PLE growth and also suggest that it is possible to selectively induce a metal-insulator transition in STO thin films fabricated under identical growth conditions, with the exception of oxygen flow rates.

We further used STEM-EELS to simultaneously understand the crystal and electronic structure of STO thin films grown under various conditions. Figure 6 presents the LAADF images and EEL spectra of representative STO films fabricated under different $P(O_2)$ and oxygen flow rates (HF and LF at $P(O_2) = 10^{-4}$ Torr). Note that the HAADF images show no significant differences among the samples (see Supplementary Fig. S5), implying that the strong contrast observed in the LAADF images can be attributed to elemental vacancies. The LAADF image in Fig. 6(a) of the STO thin film grown at $P(O_2) = 10^{-1}$ Torr shows perfect crystalline quality without any defects, equivalent to a single crystal substrate. With decreasing $P(O_2)$, the images reveal strong contrast due to local atomic dechanneling, indicative of defect formation. In particular, vacancy clusters are introduced to the films grown at low $P(O_2)$, which seem to be responsible for the observed strong tetragonality. Geometric phase analysis (GPA) specifically shows that lattice structures are locally distorted about the vacancy clusters (see Supplementary Fig. S6).[35] The LAADF images of films grown at $P(O_2) = 10^{-4}$ Torr also show distinct signs of vacancies (HF and LF for Figs. 6(c) and 6(d), respectively). Interestingly, the films with a large degree of oxygen vacancies (Figs. 6(b) and 6(d)) exhibit tightly-clustered vacancy structures compared to films with smaller quantities of oxygen vacancies (Fig. 6(c)) with more dispersed vacancy structures.

In order to investigate the evolution of vacancy-induced electronic structures in more detail, we investigated Ti $L_{2,3}$- and O $K$-edge EEL spectra of the regions indicated by rectangular boxes in the STEM images. The electronic structure of STO near the valence band is dominated by Ti-3$d$ and O-2$p$ orbitals which are susceptible to chemical bonding within the TiO$_6$ octahedra. For example, the energy separation ($\Delta$) between the two main peaks in the Ti $L_3$- or $L_2$-edges (attributed to $t_{2g}$ and $e_g$ orbital states, respectively) are indicative of octahedral crystal field splitting.[36] The film grown at $P(O_2) = 10^{-1}$ Torr shows $\Delta = $ ~2.2 eV (for $L_3$ edge for the rest of the text), corresponding to the splitting reported for stoichiometric STOs (2.2–2.3 eV). On the other hand, the film grown at $P(O_2) = 10^{-6}$ Torr shows $\Delta = $ ~1.8 eV, indicative of a substantial decrease in crystal field energy due to oxygen vacancies. The reduced number of nearest oxygen neighbors within the TiO$_6$ octahedra and the concomitant change in the Ti valence state from 4+ to 3+ resulted in a significant decrease in $\Delta$.[15,37-40] Additionally, Sr vacancy-induced tetragonal distortion could also yield similar effects in the EEL spectra. For the film grown at $P(O_2) = 10^{-4}$ Torr with (c) HF and (d) LF, we observed $\Delta$ of ~2.2 and ~1.9 eV, respectively, indicating that the LF grown film actually possessed a larger quantity of oxygen vacancies compared to the HF grown film, but smaller compared to the film grown at $P(O_2) = 10^{-6}$ Torr.

The O $K$-edge spectra also depicts the signature of oxygen vacancies based on the strong Ti-O hybridization in STO.[41] The A and B peaks denoted in Figure 6 are typically observed for STO samples in O $K$-edge spectra.[40] Peak A corresponds to the hybridized Ti-3$d$ $t_{2g}$ and O-2$p$ bands, which is known to decrease in intensity and broaden with increasing oxygen vacancies.[40] The same trends were observed in the EEL spectra of our samples, indicating that the systematic oxygen content could be controlled by changing $P(O_2)$ and/or the oxygen flow rate. Finally, peak B was attributed to the interaction between Sr-$sp$ and O-2$p$ bands. For the film grown at $P(O_2) = 10^{-1}$ Torr, a well-defined peak B was observed, again suggesting good quality of the homoepitaxial STO thin film. However, the peaks became ill-defined for other samples, indicative of Sr vacancies.

Comparison between the film grown at $P(O_2) = 10^{-1}$ (Fig. 6(a)) and $10^{-4}$ (HF) Torr (Fig. 6(c)) captures the one of the main conclusion of our study. These films showed very similar electronic structure (EEL spectra), due to the small difference in oxygen vacancy concentration. In particular, the difference in peak A of O $K$-edge spectra is small, indicating that the effect of oxygen vacancy is not dominant. Nevertheless, the film grown at $P(O_2) = 10^{-4}$ (HF) Torr presented certain defect structure in atomic level (LAADF images), which resulted in modification in the crystal structure due to the dominant role of Sr vacancies.

## Summary

In summary, we selectively controlled the formation of different types of elemental vacancies in SrTiO$_3$ homoepitaxial thin films and identified the individual role of the vacancies in inducing structural and electronic phase transitions. By tuning the plume dynamics during pulsed laser epitaxy growth, we could systematically control cation stoichiometry. On the other hand, oxygen flow rates within the vacuum chamber influenced oxygen vacancies within the STO films without substantially affecting plume dynamics, and therefore, cation stoichiometry. While it would be rather difficult to entirely separate the individual role of each type of vacancy, the formation of cation and oxygen vacancies in the STO thin films was found to mainly accompany structural (cubic to tetragonal) and electronic (insulator to metal) phase transitions, respectively. Our results suggest that efficient and selective defect engineering is achievable for complex oxides and paves the way toward exploiting defects as a means of designing physical property and functionality tailored transition metal oxide thin films and heterostructures.

## Methods

High quality homoepitaxial $Sr_xTiO_{3-\delta}$ thin films were grown on atomically flat STO single crystal substrates using PLE at 700 °C.[42] Laser (248 nm; IPEX 864, Lightmachinery, Nepean, Canada) fluence of 1.0–2.4 J/cm$^2$ with a fixed spot size (0.1 cm$^2$) and repetition rate of 5 Hz was used. The distance between the target and substrate was fixed to 65 mm. In order to systematically control the elemental vacancies in STO, films were grown under various oxygen partial pressure ($P(O_2)$) conditions ranging from $10^{-1}$ to $10^{-6}$ Torr. In addition to $P(O_2)$, we could also change the oxygen gas flow rate rather substantially, while maintaining the same $P(O_2)$. Film thicknesses were approximately 100 nm. The atomic structure, crystal orientation, and epitaxy relation of STO thin films were characterized via x-ray diffraction (XRD). The optical properties of the films were measured by using a UV-VIS spectrophotometer (200-3300 nm).

For scanning transmission electron microscopy (STEM) and electron energy loss spectroscopy (EELS), sample foils were prepared using conventional methods including mechanical thinning to ~10 $\mu$m and ion beam milling to electron transparency at an acceleration voltage of 0.5–3.5 kV using an Ar ion beam. Atomic structures were observed using a STEM (JEOL JEM-2100F, JEOL Ltd., Japan) equipped with an aberration corrector (CEOS GmbH, Heidelberg, Germany). The probe diameter of the beam was ~0.9 Å. For high (HAADF) and low angle annular dark field (LAADF) imaging, a probe convergence angle of approximately 22 mrad was used. The inner angles of the HAADF and LAADF detectors were greater than 80 and 30 mrad, respectively. Energy loss spectra were obtained at 200 kV using an EEL spectrometer (Quantum, Gatan, USA) with an energy resolution of 0.8 eV.

First-principles calculations were carried out using the screened hybrid functional of Heyd-Scuseria-Ernzerhof (HSE),[43,44] implemented with the projector augmented-wave method[45] in the VASP code[46]. The calculations for native vacancies in STO were performed, using a 3×3×3 supercell containing 135 atoms. Wavefunctions were expanded in a plane-wave basis set with an energy cutoff of 400 eV and integrations over the Brillouin zone were performed using a 2×2×2 *k*-point grid. Atomic positions were relaxed until the Hellmann-Feynman forces were reduced to less than 0.02 eV/Å.

**Acknowledgements**

This work was supported by the Basic Science Research Program through the National Research Foundation of Korea (NRF) funded by the Ministry of Science, ICT and future Planning (NRF-2014R1A2A2A01006478) and by the Ministry of Education (NRF-2013R1A1A2057523). This work was also supported by IBS-R011-D1. S.-Y. Choi, S.-D. Kim, and M. Choi were supported by Global Frontier Program through the Global Frontier Hybrid Interface Materials (GFHIM) of the National Research Foundation of Korea (NRF) funded by the Ministry of Science, ICT & Future Planning (2013M3A6B1078872). J. Hwang was supported by National Research Foundation of Korea (NRFK Grant No. 2013R1A2A2A01067629).

**Author contributions**

S.A.L and H. J conducted the experiment and analyzed the results. S.W measured the surface property. J.Y.H, H.Y and S.W.K performed structural analysis and electrical measurement. S.Y.C and S.D.K performed STEM/EELS. M.C performed theoretical calculation. S.R and J.H characterized the optical properties. S.A.L and W.S.C wrote the manuscript and all the authors reviewed the manuscript. W.S.C initiated and supervised the research.

**Additional information**

**Supplementary information** accompanies this paper at http://www.nature.com/scientificreports

**Competing financial interests:** The authors declare no competing financial interests.

**Figure**

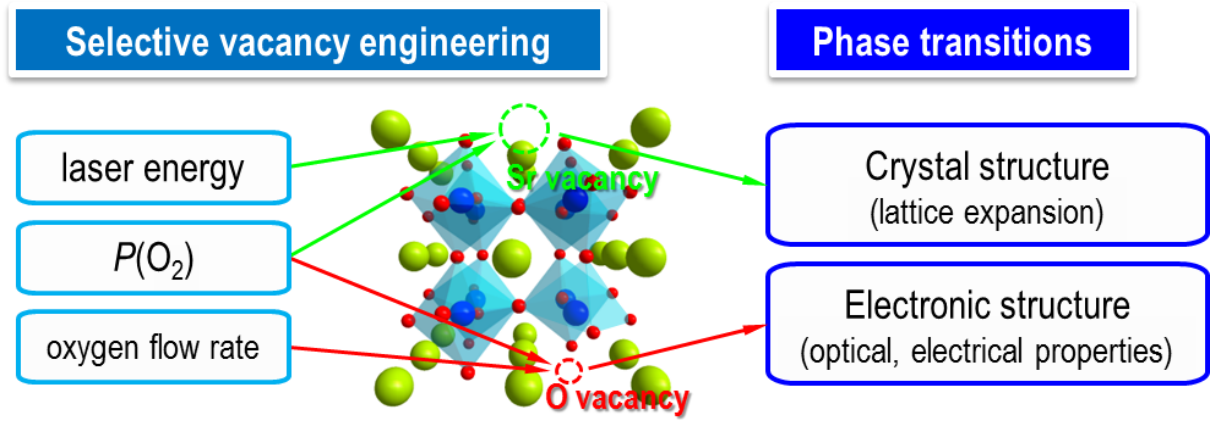

**Figure 1 |** Schematic diagram of controlling individual elemental vacancies in homoepitaxial SrTiO$_3$ thin films using pulsed laser epitaxy. Individual growth parameters in pulsed laser epitaxy (such as laser energy, partial pressure, and oxygen flow rate) result in cation and oxygen vacancies in the homoepitaxial SrTiO$_3$ thin films. Therefore, the crystal and electronic structure can be modified individually.

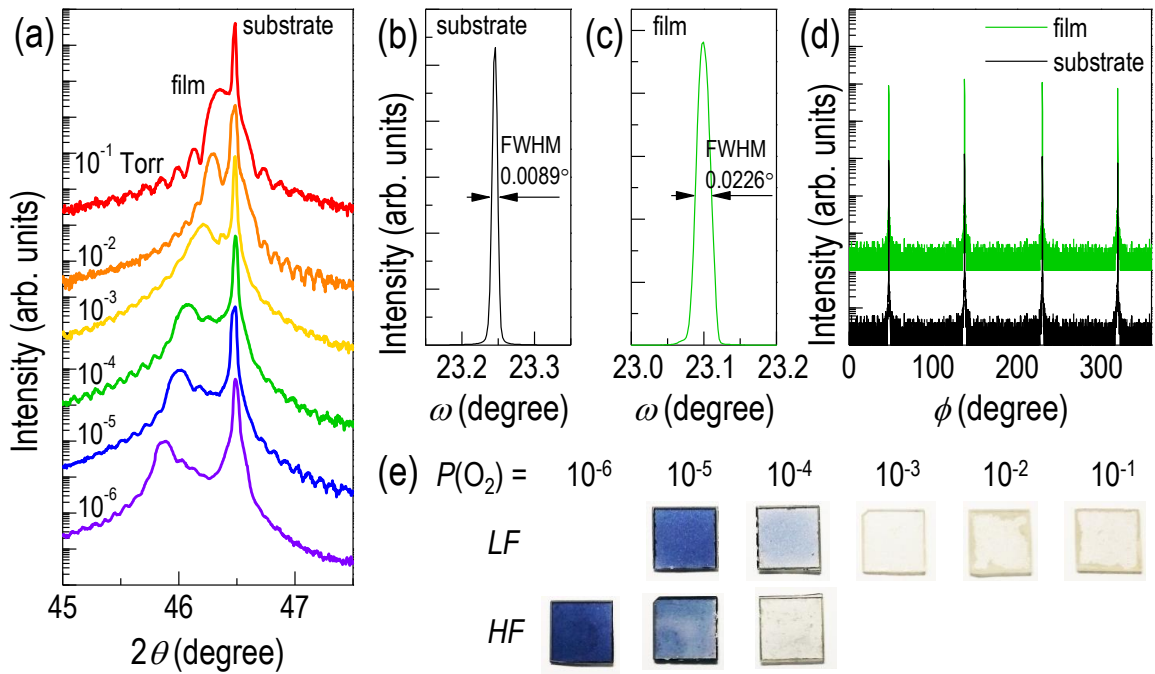

**Figure 2** | Crystal structure and colors of homoepitaxial SrTiO$_3$ thin films. (a) XRD $\theta$-$2\theta$ scans for homoepitaxial SrTiO$_3$ films grown at 200 mJ on SrTiO$_3$ substrates around the (002) Bragg plane. The films were grown at different $P$(O$_2$) from $10^{-6}$ to $10^{-1}$ Torr. Rocking curve scan of (b) the substrate and (c) the film grown at $P$(O$_2$) = $10^{-4}$ Torr. (d) Phi scans for cubic SrTiO$_3$ (024) plane (substrate) and tetragonal SrTiO$_3$ (114) plane (film grown at $P$(O$_2$) = $10^{-4}$ Torr). (e) Photographic images of homoepitaxial SrTiO$_3$ thin films. For the films grown at $10^{-4}$ and $10^{-5}$ Torr, oxygen flow rates (high flow for HF and low flow of LF) could be modified which resulted in drastic color differences.

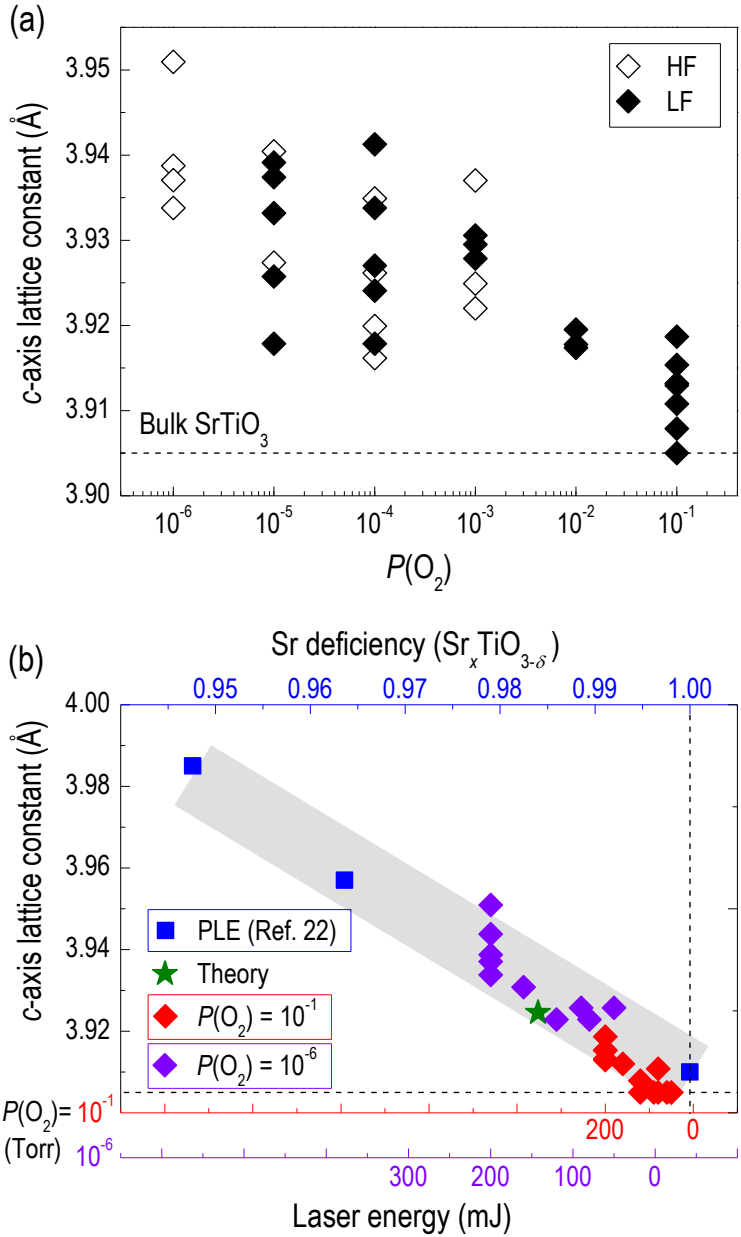

**Figure 3 |** Cation vacancy formation and *c*-axis lattice expansion of homoepitaxial SrTiO$_3$ thin films. (a) The expansion of the *c*-axis lattice constant of SrTiO$_3$ thin films grown at 200 mJ as a function of $P(O_2)$. The open (closed) symbols correspond with films grown under HF (LF). The lattice constant of bulk STO is 3.905 Å (dotted line). (b) Comparison of the *c*-axis lattice constant as a function of laser intensity and Sr deficiency (Sr$_x$TiO$_{3-\delta}$) for sets of homoepitaxial films grown by PLD (Ref. 23).

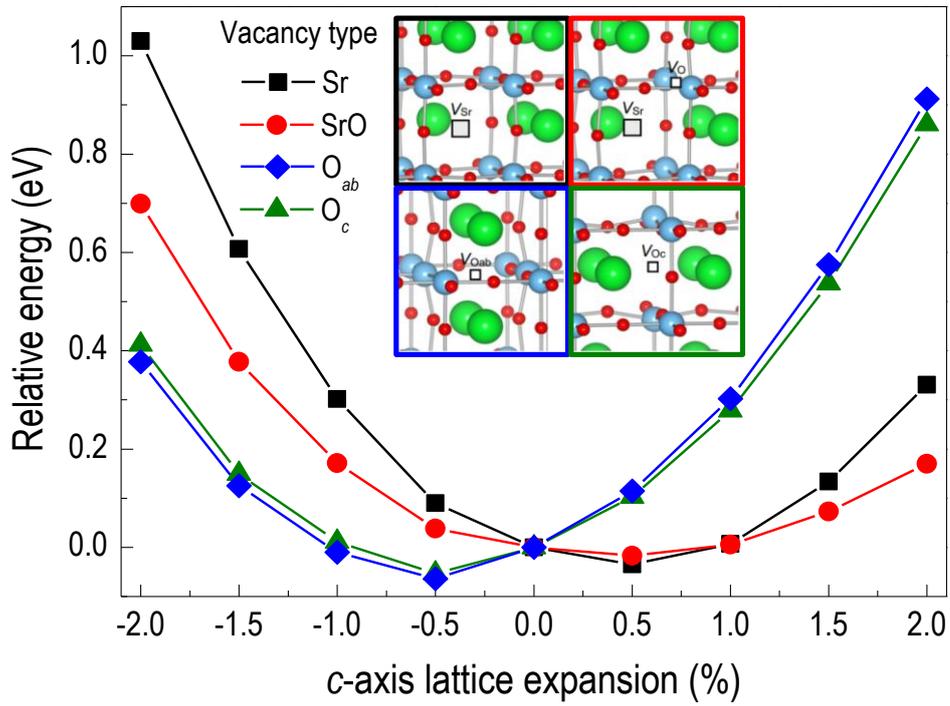

**Figure 4 |** Theoretical calculation of the relative total energy of defect-containing supercells as a function of $c$-axis lattice expansion for different elemental vacancies in SrTiO$_3$. Oxygen vacancies tend to decrease the $c$-axis lattice parameter while Sr vacancies induce $c$-axis lattice expansion. Note that Sr-O vacancies also favor larger $c$-axis lattice parameter.

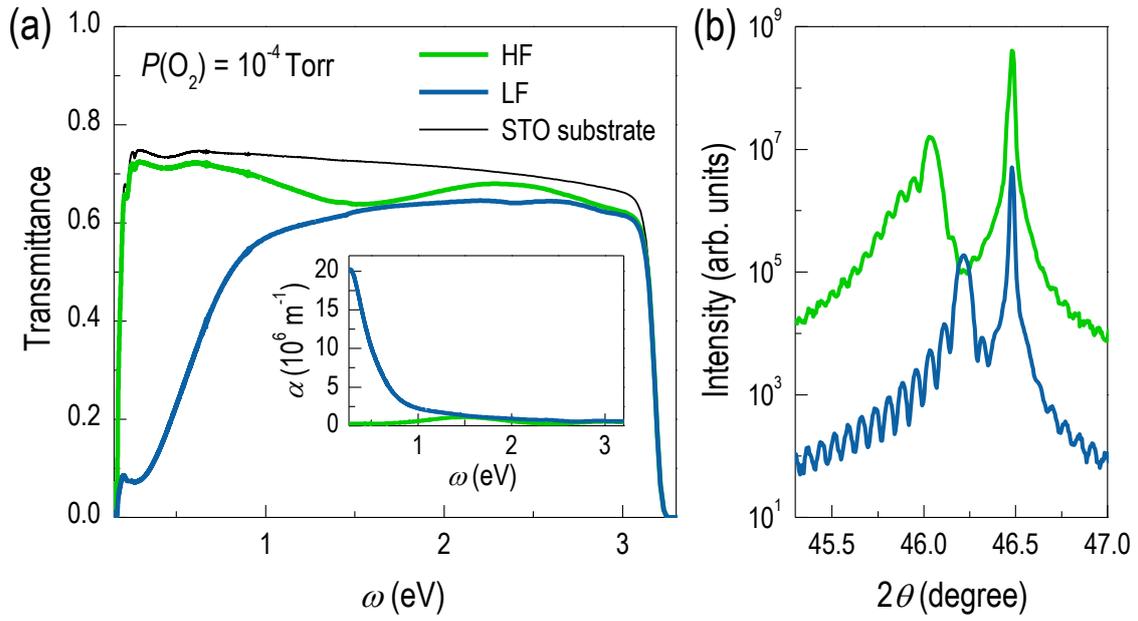

**Figure 5** | Crystal structure and optical properties of the homoepitaxial films grown under the same $P(O_2)$ but at different oxygen flow rates. (a) Optical transmittance as a function of photon energy for the same homoepitaxial films. The inset presents the absorption spectra. The HF grown film contains certainly fewer oxygen vacancies (transparent) but has a larger *c*-axis lattice parameter compared to the LF grown film. (b) XRD $\theta$-$2\theta$ scan of homoepitaxial $SrTiO_3$ films grown at $P(O_2) = 10^{-4}$ Torr with high and low oxygen flow rates (HF and LF).

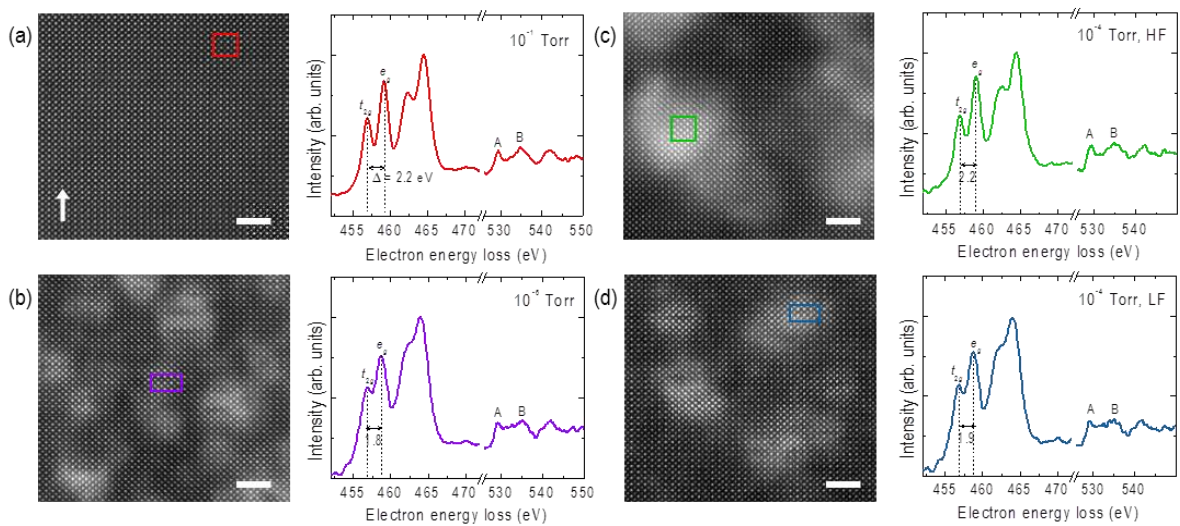

**Figure 6 |** Scanning Transmission Electron Microscopy (STEM) and corresponding electron energy loss spectroscopy (EELS) of homoepitaxial $SrTiO_3$ films. STEM low-angle annular-dark-field (LAADF) images and EEL spectra for the selected regions (squares) are shown for the homoepitaxial $SrTiO_3$ thin films grown at $P(O_2)$ = (a) $10^{-1}$ Torr, (b) $10^{-6}$ Torr, (c) $10^{-4}$ Torr with high oxygen flow rates and (d) $10^{-4}$ Torr with low oxygen flow rates. The scale bars denote 2 nm and the arrow indicates the growth direction. The peaks in the EEL spectra could be attributed to Ti-$L_3$ ($t_{2g}$, and $e_g$), Ti-$L_2$ ($t_{2g}$, and $e_g$), and O-$K$ (A and C) from the lowest energy shown. The film grown at $P(O_2) = 10^{-1}$ Torr ((a)) shows the image of perfect $SrTiO_3$ single crystal without any defects. On the other hand, the film grown at $P(O_2) = 10^{-6}$ Torr ((b)) clearly shows defective regions due to Sr and O vacancies. Films grown at $P(O_2) = 10^{-4}$ Torr ((c) and (d)) show a moderate concentration of vacancies. The EEL spectra of the $SrTiO_3$ thin films show a systematic evolution of the electronic structure with defect concentration. Ti $L$-edges reveal a systematic decrease in energy splitting ($\Delta$) between the $t_{2g}$ and $e_g$ peaks, with the introduction of oxygen vacancies. In addition, the O $K$-edges also show clear evolution of the B peak with an increase in oxygen vacancies.

**Supplementary Information**

# Phase transitions via selective elemental vacancy engineering in complex oxide thin films

Sang A Lee[1,2], Hoidong Jeong[1], Sungmin Woo[1], Jae-Yeol Hwang[3], Si-Young Choi[4], Sung-Dae Kim[4], Minseok Choi[4], Seulki Roh[1], Hosung Yu[5], Jungseek Hwang[1], Sung Wng Kim[3,5], and Woo Seok Choi[1*]

[1]Department of Physics, Sungkyunkwan University, Suwon, 16419, Korea
[2]Insitute of Basic Science, Sungkyunkwan University, Suwon, 16419, Korea
[3]Center for Integrated Nanostructure Physics, Institute for Basic Science (IBS) Sungkyunkwan University, Suwon 16419, Korea
[4]Materials Modeling and Characterization Department, Korea Institute of Materials Science, Changwon 51508, Korea
[5]Department of Energy Sciences, Sungkyunkwan University, Suwon 16419, Korea

*e-mail: choiws@skku.edu.

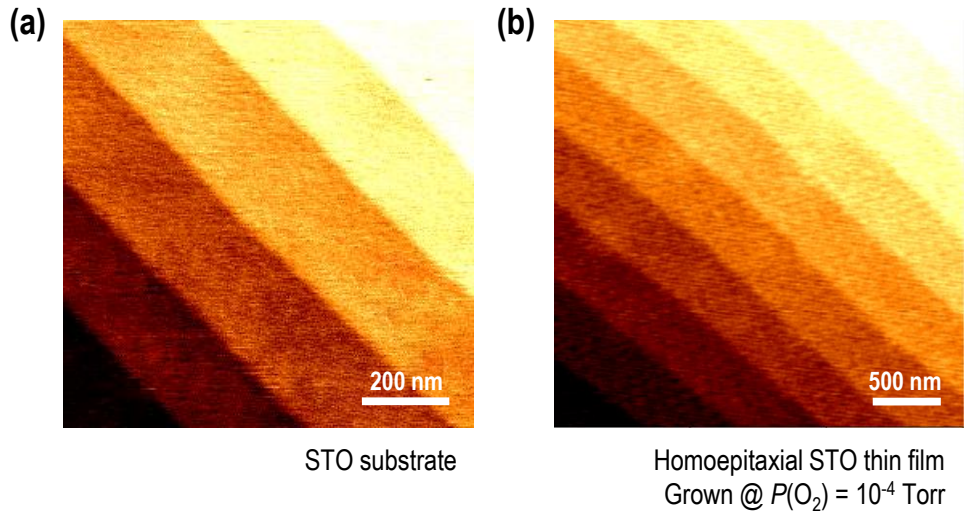

STO substrate

Homoepitaxial STO thin film
Grown @ $P(O_2) = 10^{-4}$ Torr

**Supplementary Figure S1.** Atomic Force Microscopy (AFM) topographic images of (a) SrTiO$_3$ substrate and (b) homoepitaxial SrTiO$_3$ thin film (~100 nm) grown at $P(O_2) = 10^{-4}$ Torr. AFM topographic images of the film show that the one unit cell step-and-terrace structure of the substrate is preserved.

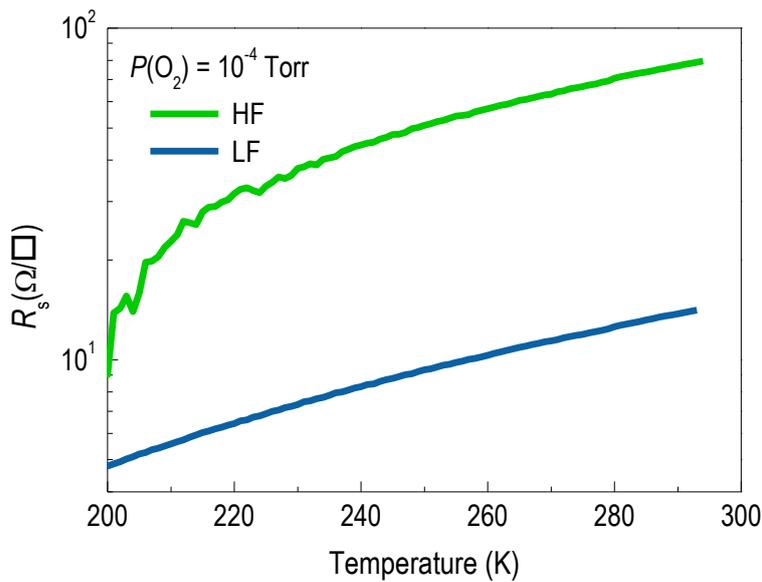

**Supplementary Figure S2.** The temperature-dependent sheet resistance, $R_s(T)$, of homoepitaxial SrTiO$_3$ films grown at $P(O_2) = 10^{-4}$ Torr with high and low oxygen flow rates (HF and LF). This result is consistent with the optical absorption spectra. The LF grown samples show more metallic nature compared to the HF grown sample due to larger oxygen vacancy concentration. We note that cation vacancies might also affect the charge transport behaviour in SrTiO$_3$ thin films, e.g., by trapping free electrons. However, in our study, the effect by the cation vacancies is much smaller than that of the oxygen vacancies which provides plenty of charge carriers.

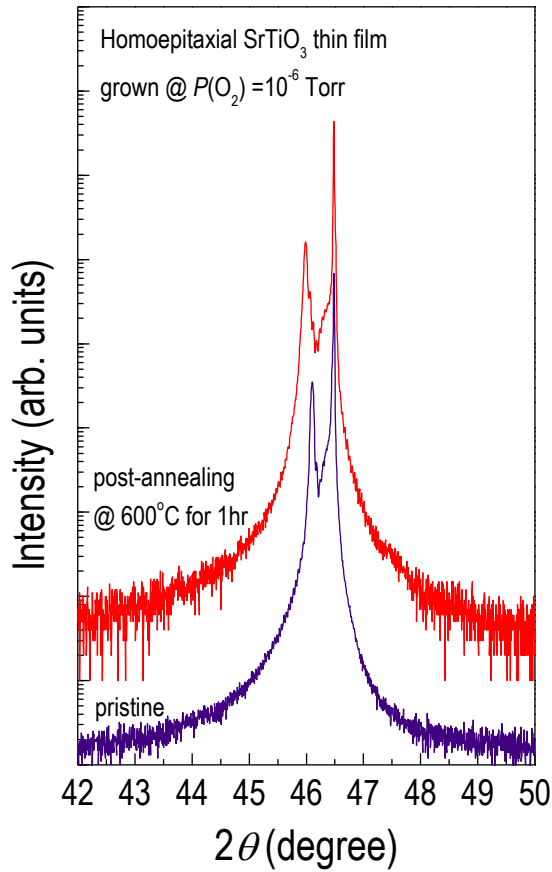

**Supplementary Figure S3.** X-ray diffraction $\theta$-$2\theta$ scans of pristine and annealed SrTiO$_3$ thin films grown at $P(O_2) = 10^{-6}$ Torr. After post-annealing in air, the *c*-axis lattice constant of the SrTiO$_3$ thin film did not change significantly, indicating that oxygen vacancies does not induce lattice expansion in SrTiO$_3$.

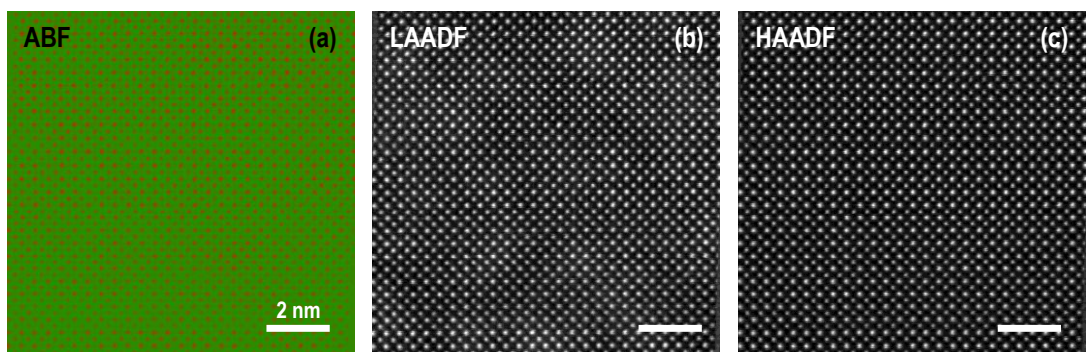

**Supplementary Figure S4.** STEM images of homoepitaxial SrTiO$_3$ thin film grown at $P(O_2) = 10^{-1}$ Torr. (a) ABF, (b) LAADF, and (c) HAADF-STEM images indicate single-crystalline thin film without any defects expected for stoichiometric SrTiO$_3$.

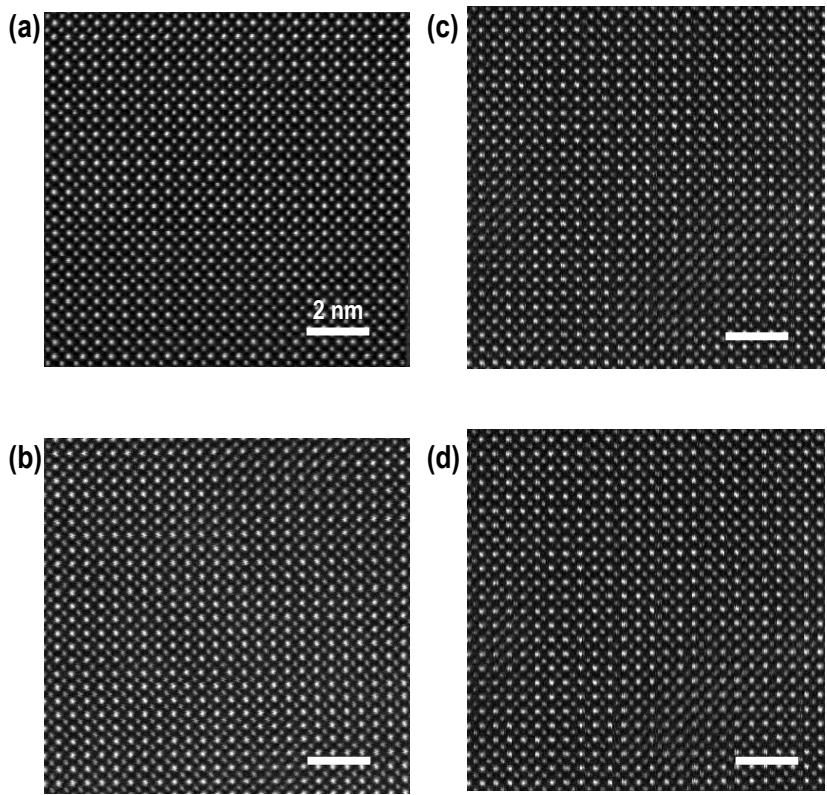

**Supplementary Figure S5.** HAADF-STEM images of homoepitaxial SrTiO$_3$ thin film. The HAADF images show no significant difference among the SrTiO$_3$ thin film grown at $P(O_2)$ = (a) $10^{-1}$ Torr, (b) $10^{-6}$ Torr, (c) $10^{-4}$ Torr with HF, and (d) $10^{-4}$ Torr with LF.

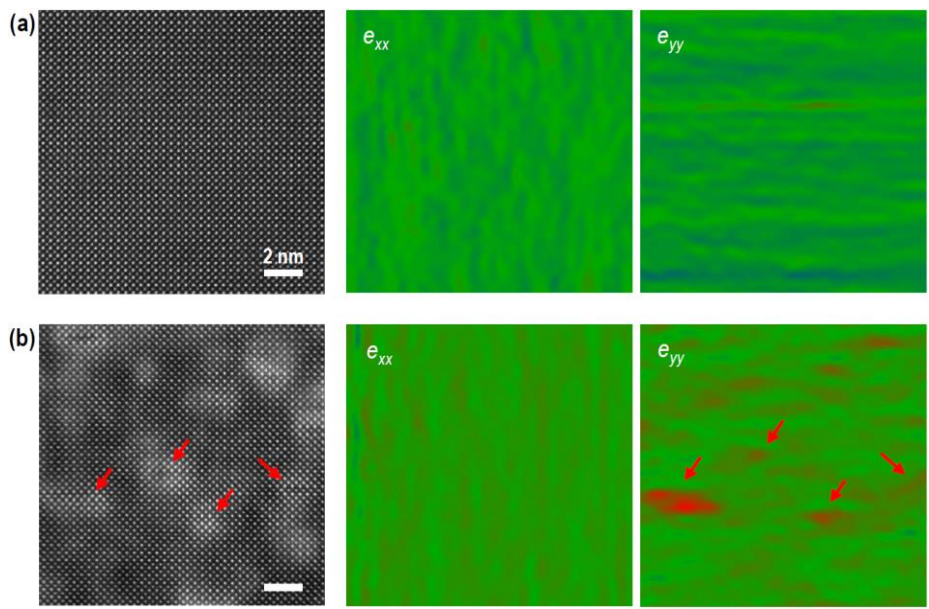

**Supplementary Figure S6**. Geometric phase analyses (GPA) of homoepitaxial SrTiO$_3$ thin films. LAADF-STEM and GPA images are shown for the corresponding spot for the films grown at $P(O_2)$ = (a) $10^{-1}$ and (b) $10^{-6}$ Torr. The film grown at low pressure shows substantial local lattice distortion due to vacancy cluster formation.

28